\documentclass[apj]{emulateapj}
\newcommand{\jcap}{JCAP}

\usepackage{natbib}

\newcommand{\mpch}{~\>h^{-1}{\rm {Mpc}}}
\newcommand{\degree}{^{\circ}}
\newcommand{\ud}{\,\rm{d}}

\newcommand{\thetaa}{\theta_{\mathrm{A}}}
\newcommand{\smh}{\>h^{-1}\mathrm{M}_{\odot}}

\newcommand{\rmag}{\>^{0.1}{\rm M}_r-5\log h}
\newcommand{\gmr}{\>^{0.1}{(g-r)}}
\newcommand{\thetam}{\langle \theta \rangle}

\usepackage{color}

\begin{document}

\title{BSG alignment of SDSS galaxy groups}


\author{Zhigang Li\altaffilmark{1,2,*},
        Yougang Wang\altaffilmark{1,$\wr$},
        Xiaohu Yang\altaffilmark{3,4,$\dagger$},
        Xuelei Chen\altaffilmark{1,5,$\ddagger$},
        Lizhi Xie\altaffilmark{2,6,$\S$},
        Xin Wang\altaffilmark{7,$\flat$}  }

\altaffiltext{1}{Key Laboratory of Optical astronomy, National Astronomical
  Observatories,Chinese Academy of Science, Beijing 100012, China}

\altaffiltext{2}{Graduate University of Chinese Academy of Sciences, Beijing
  100049, China}

\altaffiltext{3}{Center for Astronomy and Astrophysics, Shanghai Jiao Tong
  University, Shanghai 200240, China}

\altaffiltext{4}{Key Laboratory for Research in Galaxies and Cosmology,Shang
  Astronomical Observatory, Nandan Road 80, Shanghai, 200030, China}

\altaffiltext{5}{Center of High Energy Physics, Peking University, Beijing
  100871, China}

\altaffiltext{6}{National Astronomical Observatories,Chinese Academy of
  Science, Beijing 100012, China}

\altaffiltext{7}{Department of Physics and Astronomy, Johns Hopkins
  University, Baltimore, MD 21218, US}

\altaffiltext{*}{zgli@bao.ac.cn}
\altaffiltext{$\wr$}{wangygcluster@gmail.com}
\altaffiltext{$\dagger$}{xhyang@shao.ac.cn}
\altaffiltext{$\ddagger$}{xuelei@cosmology.bao.ac.cn}
\altaffiltext{$\S$}{lzxie@bao.ac.cn}
\altaffiltext{$\flat$}{wangxin@pha.jhu.edu}

\begin{abstract}
We study the alignment signal  between the distribution of brightest satellite
galaxies  (BSGs) and  the major  axis of  their host  groups using  SDSS group
catalog constructed by Yang et al. (2007).  After correcting for the effect of
group  ellipticity, a  statistically significant  ($\sim  5\sigma$) major-axis
alignment is detected and the alignment  angle is found to be $43\degree.0 \pm
0.4$.  More  massive and richer  groups show stronger BSG  alignment. The
  BSG alignment  around blue  BCGs is slightly  stronger than that  around red
  BCGs. And red  BSGs have much stronger major-axis  alignment than blue BSGs.
  Unlike BSGs,  other satellites do  not show very significant  alignment with
  group major  axis.  We  further explore the  BSG alignment  in semi-analytic
  model  (SAM) constructed  by  Guo et  al.   (2011).  We  found general  good
  agreement with  observations: BSGs in  SAM show strong  major-axis alignment
  which depends  on group mass and  richness in the same  way as observations;
  and  none  of  other   satellites  exhibit  prominent  alignment.   However,
  discrepancy also exists in that the SAM shows opposite BSG color dependence,
  which is  most probably  induced by the  missing of large  scale environment
  ingredient in SAM. {\bf The combination of} two popular scenarios can explain 
  the detected BSG alignment.  The first one: satellites merged into the group
  preferentially along  the surrounding  filaments, which is  strongly aligned
  with the  major axis of  the group.  The  second one: BSGs enter  their host
  group  more  recently  than   other  satellites,  then  will  preserve  more
  information about the assembling history and so the major-axis alignment. In
  SAM, we  found positive evidence  for the second  scenario by the  fact that
  BSGs  merged  into   groups  statistically  more  recently  than  other
  satellites.  We also found that most  of BSGs ($80\%$) were BCGs before they
  merged into groups  and earlier merged BSGs tend to be  closer to their BCGs
  than other  BSGs. On  the other hand,  although is opposite in SAM,  the BSG
  color dependence in observation might indicate the first scenario as well.

\end{abstract}


\maketitle

\section{Introduction}      
\label{introduction}

In the current understanding of the hierarchical structure formation scenario
of cold dark matter (CDM), small dark matter halos form first from the
anisotropic collapse of overdensities in the mass distribution, while larger
halos grow by accreting surrounding material and/or by merging with other
halos. It is well believed that satellites in clusters or groups trace the
distribution of dark matter halos. Therefore, it is popular to study the shape
or mass distribution of dark matter halos using the distribution or kinematics
of satellite galaxies (\citealt{2004MNRAS.352.1323P, 2006ApJ...650..770P,
  2008MNRAS.385.1511W, 2010MNRAS.407..581R, 2009MNRAS.392..917M,
  2009MNRAS.392..801M, 2011MNRAS.410..210M, 2010MNRAS.407....2D,
  2012JCAP...05..030S, 2012arXiv1201.1296G}).  Among others, a power tool is
to measure the alignment signals between the distribution of satellites and
the reference directions, e.g.  the major axes of central galaxies, groups or
halos, etc.

Intrinsic alignment of galaxies relative to their host dark matter halos
encodes plenty of information of galaxy formation and evolution history and
their surrounding filamentary structures (\citealt{2007ApJ...655L...5A,
  2005ApJ...627..647B, 2008MNRAS.389.1266L, 2009JCAP...06..009S,
  2011MNRAS.416.1377V, 2012MNRAS.421L.137L}).  Observationally, three types of
alignment have been extensively studied in the literatures: (1) the satellite
alignment, between the distribution of all satellite galaxies and the major
axis of their central galaxy (\citealt{1969ArA.....5..305H,
  1979MNRAS.187..287S, 1982MNRAS.198..605M, 1997ApJ...478L..53Z,
  2004MNRAS.348.1236S, 2009MNRAS.395.1184S, 2005ApJ...628L.101B,
  2006MNRAS.369.1293Y, 2010ApJ...709.1321A, 2007MNRAS.376L..43A,
  2007ApJ...662L..71F, 2009RAA.....9...41F, 2008MNRAS.385.1511W,
  2008MNRAS.387.1199S, 2010ApJ...718..762W, 2011ApJ...731...44N,
  2011ApJ...740...39H}); (2) the radial alignment, between the orientation of
satellite galaxies and the central-satellite connection line or the
orientation of brightest central galaxies (BCGs)
(\citealt{2005ApJ...627L..21P, 2006ApJ...644L..25A, 2007ApJ...662L..71F,
  2011ApJ...740...39H}); (3) the alignment between the shape of groups and the
large scale structure (\citealt{2008MNRAS.389.1127P, 2011MNRAS.414.2029P,
  2009RAA.....9...41F, 2009ApJ...703..951W}). It is notable that the
measurements of alignment between satellites and their host galaxies have a
long and confused history. Some studies reported a minor axis alignment, while
most of others preferred a major axis alignment.  We refer those who are
interested in this history to \cite{2006MNRAS.369.1293Y} and
\cite{2008MNRAS.385.1511W} for summaries. Now it is clear that the satellites
preferentially lie along the major axis of host galaxies, i.e. a major axis
alignment.  Moreover, the alignment depends on the properties of the host
halos, such as halo mass, color of central galaxies and satellites.  For
example, more massive halos usually show stronger alignment, and red BCGs and
satellites show stronger alignment than the blue ones. These are consistent
with the simulation results, where the distribution of subhalos is aligned
with major axis of their host halos and more massive halos show a stronger
alignment signal (\citealt{2005MNRAS.363..146L, 2005MNRAS.364..424W,
  2005ApJ...629..219Z, 2007MNRAS.378.1531K, 2008ApJ...675..146F,
  2008MNRAS.386L..52K, 2008MNRAS.388L..34K, 2010MNRAS.405.1119K}).

Many dynamical processes can contribute to the alignment of satellites
associated with groups, which can be roughly divided into two classes - large
scale environment and the impact of group potential
(\citealt{2000AJ....119.2248H, 2010ApJ...718..762W}).  The filamentary
structures surrounding the group may readjust the satellites pointing toward
the group. Additionally, the anisotropic accretion along these filaments will
make the satellites preferentially distributed along the group major axis.
While, phase-mixing effect and relaxation will smear the memory of assembling
history of satellites when they merged into the potential well of the
groups. In this respect, more recently accreted or merged satellites contain
more information about the assembling history and should show stronger
alignment with the major axis of the host halo than other satellites.

It is expected that the brightest satellite galaxies (BSGs) enter the host
group mainly through major mergers and play a determinant role in shaping the
host group. If the BSGs enter the host halos statistically more recently, they
will preserve the most information about the merger history. In this case,
BSGs will be good tracers of the structure formation of their host groups.

In this paper, we investigate the alignment of BSGs with the major axis of
their host groups using the SDSS DR4 galaxy group catalog constructed by
\cite{2007ApJ...671..153Y}. Different from the previous satellite alignment
measurements, we focus on the alignment signal between the distribution of
BSGs and the major axis of their host halos, not the major axis of the central
galaxies. While the major axis of the central galaxy itself is misaligned with
the projected major axis of the host, e.g. at the level of $\sim 23\degree$
\citep{2008MNRAS.385.1511W}. Our observational measurements are then compared
with the BSG alignment in semi-analytic model (SAM) of galaxy formation catalog
constructed by \cite{2011MNRAS.413..101G}.  We also investigate the
distribution of the time when BSGs merged and compare them with those for
other satellites in SAM.

This paper is organized as following. In \S 2, we introduce the SDSS group
catalog and SAM model briefly. The alignment estimator we used is described in
\S 3. Then we show the results of BSG alignment in SDSS groups in \S 4 and in
SAM in \S 5. In \S 5 we also show the distribution of BSG merger time. Finally
a brief summary and discussions are given in \S 6.

\section{Data} \label{data}

\subsection{Groups of galaxies}

Our analysis is based on the SDSS DR4 galaxy group catalog, which is
constructed by \cite{2007ApJ...671..153Y} (hereafter Y07) using an adaptive
halo-based group finder (\citealt{2005MNRAS.356.1293Y}).  For a full
description of the catalog the reader is referred to their paper, but the
major part of the catalog is presented briefly here.

The base galaxy catalog is the New York University Value-Added Galaxy
Catalogue of SDSS-DR4 (NYU-VAGC; \citealt{2005AJ....129.2562B}).  As described
in Y07, three group samples were constructed: Sample I, which only uses the
362,356 galaxies with measured $r$-band magnitudes and redshifts from the
SDSS, Sample II which also includes 7,091 galaxies with SDSS $r$-band
magnitudes and redshifts taken from other surveys, and Sample III which
includes an additional 38672 galaxies that fail to get redshifts due to fiber
collisions and are assigned the redshifts of their nearest neighbors
(\citealt{2005ApJ...630....1Z}). Our analysis is based on Sample III.

All magnitudes are extinction corrected (\citealt{1998ApJ...500..525S}) and
k-corrected (\citealt{2003AJ....125.2348B}) and evolved to rest-frame
magnitudes at $z=0.1$ using the evolving luminosity model of
\cite{2003ApJ...592..819B}. Stellar mass of galaxies are estimated according
to the fitting formula of \cite{2003ApJS..149..289B}.  The galaxies are
assigned red or blue color according to their bi-normal distribution in the
$\gmr$ color (\citealt{2004ApJ...600..681B, 2006MNRAS.368...21L}). The galaxy
is red if $\gmr > 1.022-0.0651x-0.00311x^2$, where $x=\rmag+23$, and blue
otherwise \citep{2008ApJ...676..248Y}. The dark matter halo mass of group is
estimated based on the ranking of the characteristic group stellar
mass. Survey edge effects have been taken into account by removing groups
($\sim 1.6\%$) that suffer severely from edge effects. Since the redshifts
assigned according to the nearest neighbors are not very reliable, we discard
groups whose brightest or second brightest galaxies have assigned
redshifts. The total number of groups in the resulted catalog is 8,513.

The brightest galaxy in the group is taken as the central galaxy and all
others are satellite galaxies. In addition, we also considered the most
massive galaxy (in terms of stellar mass) as the central galaxy (MCG). As we
have checked, the difference of the alignment signals between the luminosity
(e.g. BCG) and stellar mass (e.g. MCG) indicators is too small to be
noted.

\subsection{Semi-analytic model}

The semi-analytic catalog of galaxy population used in this paper was
constructed by \cite{2011MNRAS.413..101G} based on the combination of
Millennium (MS) and MS-II simulation. The MS and MS-II simulation are run by
the Virgo Consortium (\citealt{2009MNRAS.398.1150B}) and both contain $2160^3$
particles. They adopted the same $\Lambda\mathrm{CDM}$ cosmology with
parameters $\Omega_\mathrm{tot}=1$, $\Omega_\mathrm{m}=0.25$,
$\Omega_\mathrm{b}=0.045$, $\Omega_\mathrm{\Lambda}=0.75$, $h=0.73$,
$\sigma_8=0.9$ and $n_s=1$.  MS has a box size of $500\mpch$ and each particle
has mass $8.6\times10^8\smh$, while the box size of MS-II is $100\mpch$ and
particle mass is $6.9\times10^6\smh$. Dark matter halos in MS and MS-II
simulation are found by a friend-of-friend (FOF) method with linking length
being 0.2 times of the mean separation of particles
(\citealt{1985ApJ...292..371D}).  The SUBFIND algorithm
(\citealt{2001MNRAS.328..726S}) was applied to each group to identify all
bound substructures (subhalos).

The galaxy formation models are implemented by allowing the galaxies to grow
at the potential minima of the halos and subhalos in simulation.  Each FOF
group contains a central galaxy at the potential minimum of its main
subhalo. Other satellite galaxies may reside at the potential minima of
satellite subhalos, or be ``orphans'' which no longer correspond to any
subhalos. Various semi-analytic effects are considered, such as tidal torque,
dynamical friction, star formation, gastrophysics, supernova and active
galactic nucleus feedback and galaxy mergers and etc. We refer those who are
interested to \cite{2011MNRAS.413..101G} for more details about the SAM catalog.
Here galaxies are also assigned red and blue colors according to the
  $(g-r)$ color bi-normal distributions within these SAM galaxies.

\section{Method}  \label{method}

For a given galaxy group, the principal axis and its orientation on
the sky can be calculated from the inertia tensor
\begin{equation}
X_{ij} = \sum_{n=1}^{N_\mathrm{mem}} x_{i,n} x_{j,n}/r^2; ~~~
i,j=1,2
\end{equation}
where $N_\mathrm{mem}$ is the number of group members, $(x_{i,n}, x_{j,n})$
are the projected coordinates (with BCG as origin) of the \emph{n}th satellite
galaxy, and $r$ is the distance between BCG and group members.  The $1/r^2$
weight is used to avoid overweighting on the group members at outskirts.  The
semi-major and semi-minor axes of the ellipse, $L_a$ and $L_b$, can be derived
by solving the eigenvalue problem,
\begin{equation}
\left| \begin{array}{cc}
X_{11}-L^2 & X_{12} \\
X_{12}     & X_{22}-L^2 \\
\end{array} \right|
= 0 \; \; \mbox{.}
\end{equation}
The direction of the major axis is given by the eigenvector ${\textbf{r}} =
[1,(L_a^2-X_{11})/X_{12}]$, while the ellipticity, $\epsilon$, and the axis
ratio, $\eta$, are
\begin{equation}
\epsilon=1-L_b/L_a,~~~\mbox{and}~~~~ \eta=L_b/L_a\,.
\end{equation}

We quantify the distribution of BSGs relative to the group major axes 
by the distribution function,
\begin{equation}
P(\theta)  = \frac {N(\theta)}{N_\mathrm{tot}}
\end{equation}
where $\theta$ is the angle between the group major axis and the BSG-BCG
connection line. The angle $\theta$ is constrained in the range
$0\degree\leq\theta\leq90\degree$, where $\theta=0\degree (90\degree)$
indicates that the BSGs lie along the major (minor) axis of the host group.

The strength of alignment can be quantified by the average of $\theta$,
\begin{equation}
\thetam = \int P(\theta) \theta \ud \theta
\end{equation}
Then $\thetam < 45\degree$ indicates the alignment of BSGs with the group
major axes, otherwise minor axes. The alignment is stronger if $\thetam$ is
further from $45\degree$ and there is no alignment at all if $\thetam =
45\degree$. This estimator works well only if the groups are spherical. For
elliptical groups, the alignment angle defined as this will generally depart
from $45\degree$, even if the BSGs are randomly distributed. To consider the
effect of group shape, one can generate mock samples in which the BSGs are
randomly distributed according to the projected two dimensional axes of
groups. To get the true alignment signals of the BSGs with respect to other
satellite galaxies in the non-spherical groups, we define
$P_{\mathrm{A}}(\theta) = P(\theta) - P_{\mathrm{mock}} (\theta) + w$ with the
normalization factor $w$, which can be calculated by $\int
P_{\mathrm{A}}(\theta) d\theta = 1$, and the alignment angle $\thetaa = \int
P_{\mathrm{A}}(\theta) \theta \ud \theta = \thetam_\mathrm{data} -
\thetam_\mathrm{mock} + 45\degree$. In practice, we construct mock catalog by
randomly selecting the satellites as the BSGs in the groups, which reflect the
distribution of all satellites. The ellipticity effect and shot noise are
included automatically. The distribution functions for these mock samples are
measured and averaged to get the mean angle,
$\thetam_{\mathrm{mock}}$, which are merely contributed by the shape of
groups.  We generated 1000 mock samples to calculate the errors of
distribution function and alignment angle.

\section{Alignment of BSGs in SDSS groups} \label{result}

\begin{figure}
\resizebox{\hsize}{!}{\includegraphics{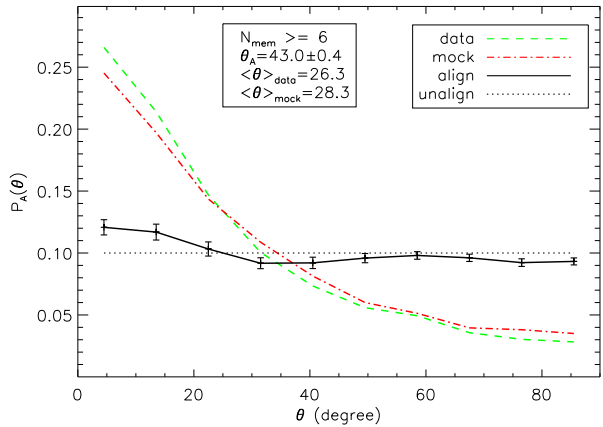}}
\caption{ Alignment of BSG and group major axis. The green dashed line shows
  the normalized distribution function of the angle between BSG-BCG connection
  line and group major axis. While the mean distribution function of mock
  samples is shown as red dot-dashed line. The black solid line with error
  bars is the true alignment distribution function. }
\label{full}
\end{figure}

In this section, we study the alignment signal between BSG-BCG connection line
and the major axis of the host group. Firstly we measure the distribution
function and alignment signal for the full sample in which each group has at
least 6 members.  The results are shown in FIG. ~\ref{full}. The true
alignment distribution function, $P_{\mathrm{A}}(\theta)$, is shown as black
solid line with error bars come from 1000 mock samples. A significant group
major axis alignment of BSGs is detected. This can be seen from the shape of
the alignment distribution function, and the resulting alignment angle which
is $\theta_{\mathrm{A}}=43\degree.0\pm0.4$ and smaller than $45\degree$.

As a  comparison, the distribution  functions, $P(\theta)$, for data  and mock
sample  are also  shown  in FIG.   \ref{full}  as green  dashed  line and  red
dot-dashed line  respectively. Although the distribution function  for data is
much      steeper       and      the      corresponding       mean      angle,
$\thetam_\mathrm{data}=26\degree.3$,  is much  smaller,  when the  ellipticity
effect  is  subtracted  using  mock  sample,  the  true  alignment  signal  is
reconstructed.

In the following we explore the dependence of BSG alignment on various group
properties, including halo mass of groups in \S \ref{secmass}, BCG and BSG
color in \S \ref{seccolor} and group richness in \S \ref{secrich}.  To answer
the question whether the BSGs are different from other satellites, we compare
the alignment signal of BSG with that of other satellites in \S \ref{secsat}.

\subsection{Dependence on group mass} \label{secmass}

\begin{figure}
\resizebox{\hsize}{!}{\includegraphics{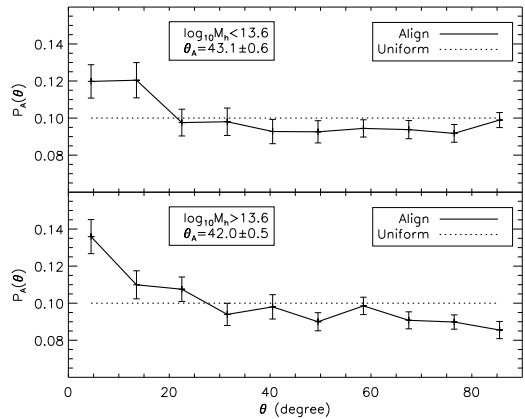}}
\caption{Dependence of BSG alignment on group mass. Upper panel shows the
  results for low mass groups, and lower panel for high mass ones.  }
\label{mass}
\end{figure}

Previous alignment studies in groups, which measure the alignment between the
distributions of satellites and the major axes of central galaxies, have shown
clear mass dependence, i.e., the alignment is stronger in more massive groups
(\citealt{2006MNRAS.369.1293Y, 2008MNRAS.385.1511W} and references there in).
In this section, we explore whether this mass dependence exists for BSG
alignment. Limited by the size of SDSS group catalog, we divide them in to 2 
subsamples according to their halo mass. Low mass sample contains groups 
with mass $\log_{10}M_\mathrm{h} < 13.6$ and high mass sample contains others. 
Where $M_\mathrm{h}$ is halo mass with unit $\smh$. The BSG alignment for them 
are shown in FIG. \ref{mass}.  

\subsection{Dependence on BCG and BSG color} \label{seccolor}

\begin{figure}
\resizebox{\hsize}{!}{\includegraphics{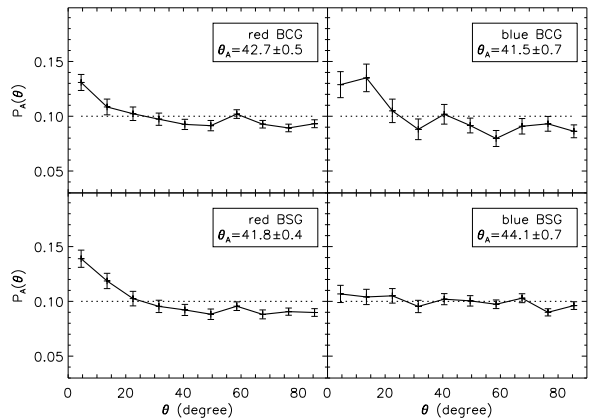}}
\caption{BSG alignment dependence on BCG and BSG color. The upper panels show
  BSG alignment for groups with red (left) and blue (right) BCGs, while the
  lower panels for groups with red (left) and blue (right) BSGs
  respectively. }
\label{color}
\end{figure}

In this section we study how the BSG alignment depends on the color of BCG and
BSG. Following \cite{2006MNRAS.369.1293Y} and \cite{2008MNRAS.385.1511W}, we
divide the groups into four subsamples according to the color of BCGs and BSGs
and measure the alignment signal of BSGs for them. The results are shown in
FIG. \ref{color}. BSGs in groups with either red BCGs or blue BCGs show
significant alignment with major axis of their host groups, with a little bit
stronger alignment in groups with blue BCGs. On the other hand, the
  results do show quite prominent BSG color dependence in that red BSGs show
  much stronger alignment with major axis of their host groups.

\subsection{Effect of group richness} \label{secrich}

\begin{figure}
\resizebox{\hsize}{!}{\includegraphics{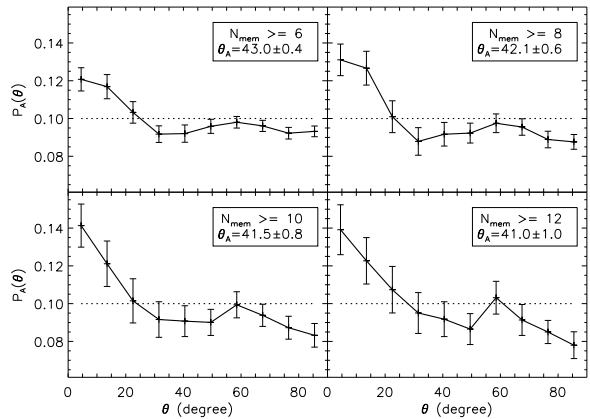}}
\caption{BSG alignment dependence of group richness. The results for groups
  with at least 6 members are shown in upper-left panel, 8 members in
  upper-right, 10 members in lower-left and 12 members in lower-right panel
  respectively. }
    
\label{richness}
\end{figure}

To study the effect of group richness, we construct subsamples for groups with
different richness and compare the alignment signal for them.  The results for
groups with at least 6, 8, 10 and 12 members are shown in FIG. \ref{richness}.
Although the measurement errors for rich groups are relatively larger, the
trend that BSGs in richer groups are more strongly aligned with the major axis
of their host groups, is clearly shown.

\subsection{Comparison with other satellites} \label{secsat}

\begin{figure}
\resizebox{\hsize}{!}{\includegraphics{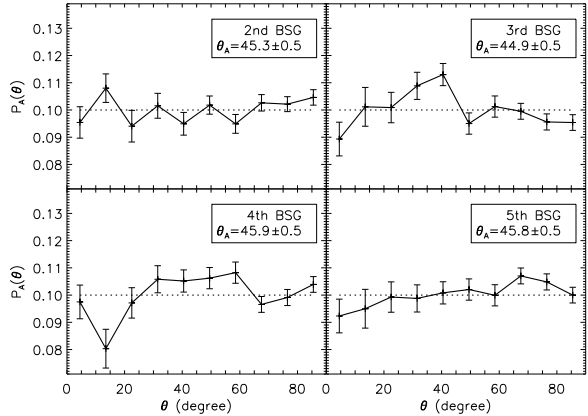}}
\caption{Alignment of other satellites for groups with more than 
6 members. Upper-left panel shows results for the 2nd BSG, upper-right
and lower panels for 3rd, 4th (left) and 5th (right) brightest satellites
respectively. }
\label{sat6}
\end{figure}

Finally, we study the alignment of other satellites with respect to the group
major axis to see whether the BSG is special in this respect. In
  FIG. \ref{sat6}, we show the results for the second to fifth brightest
  satellites. Interestingly, all of them do not show any significant
  alignment, with alignment angles consistent with $45\degree$ at 2-$\sigma$
  level.

\section{BSG Alignment and Merger time in SAM} \label{secsam}

In this section, we study the BSG alignment with group major axis
in SAM galaxy formation model constructed by \cite{2011MNRAS.413..101G}.
To explore the origin of the BSG alignment, we also measure the
time when BSGs merged into their host halos.

\subsection{BSG Alignment}

\begin{figure}
\resizebox{\hsize}{!}{\includegraphics{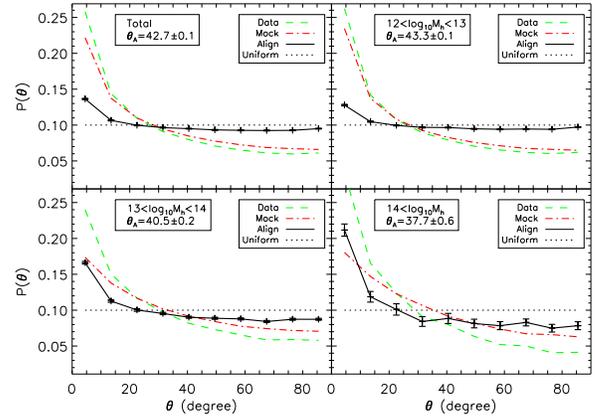}} 
\caption{BSG alignment with
  group major axis and its dependence on group halo mass in SAM galaxy
  formation model. }
 \label{sam}
\end{figure}

In SAM galaxy formation model, we select groups which have at least 6 members
with stellar mass greater than $10^8 \smh$. The group major axis is estimated
using the method discussed in \S \ref{method}. BSGs are selected by their rank
of stellar mass in each group. The BSG alignment with group major axis in SAM
and their dependence on group halo mass are shown in FIG. \ref{sam}. Clearly,
BSGs show major-axis alignment and the alignment angle is $42.7\degree
\pm0.1$.  Furthermore, more massive groups show stronger BSG alignment.
  Especially in massive groups with $\log M_h > 14.0$, the alignment angle
  $37.7\degree\pm 0.6$ is much smaller than that obtained in SDSS observation. This
  can be understood in several ways. The first possibility is that as
  discussed in \cite{2007MNRAS.378.1531K}, the various observational selection
  effects the group catalog suffered from, such as interlopers (groups members
  do not actually belong to the same group) and incompleteness, tend to blur
  the true alignment signal. Additionally, the hydrodynamical simulations,
  including baryon physics, have indicated that the condensation of baryons
  tend to make halo more spherical or axisymmetric and cause the
  misalignment between observations and N-body simulations
  (\citealt{2006ApJ...651..636L, 2008ApJ...681.1076D, 2010MNRAS.402..776P,
    2010MNRAS.406..922T,  2012MNRAS.422.1863B, 2013MNRAS.tmp..566B,
    2012ApJ...748...54Z} and references therein).

\begin{figure}
\resizebox{\hsize}{!}{\includegraphics{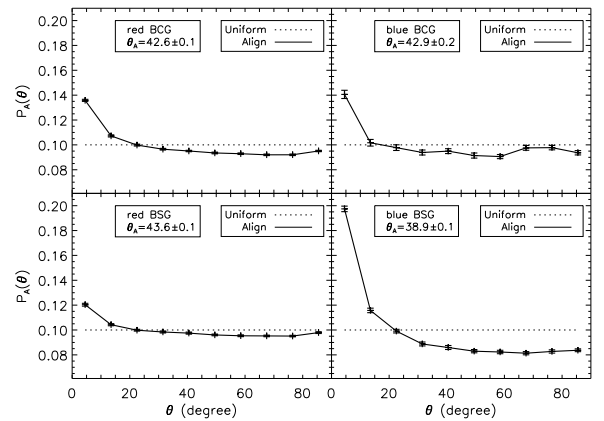}}
\caption{Dependence of BSG alignment on color of BCGs and BSGs in
  SAM.}
 \label{samcolor}
\end{figure}

  In FIG. \ref{samcolor}, we show the color dependence of the alignment
  signals. First, for BCGs of different colors, the BSG alignments are quite
  similar to observations and no significant color dependence is
  found. Second, for BSGs of different colors, quite similar to the
  observations, there exhibit prominent color dependence, however, in {\it
    opposite} direction. This discrepancy is quite interesting. As the
  stronger alignment of the BSGs with respect to the group major axes
  indicates the more recent accretion along the filaments, both results in
  SDSS observation and SAM are rather expected.  In observation, the real
  Universe, these BSGs before their accretion into groups, are preferentially
  located in filaments where the gas temperature might be quite high due to
  shock heating, etc., and are thus quenched and red. On the other hand, in
  SAM, although these BSGs before accretion might also be preferentially
  located in the filaments, since SAM only make use of the halo merger trees
  where such kind of filament environment is not taken into account, and thus
  these BSGs (BCGs before accretion) might still be star forming and blue.  
 
\begin{figure}
\resizebox{\hsize}{!}{\includegraphics{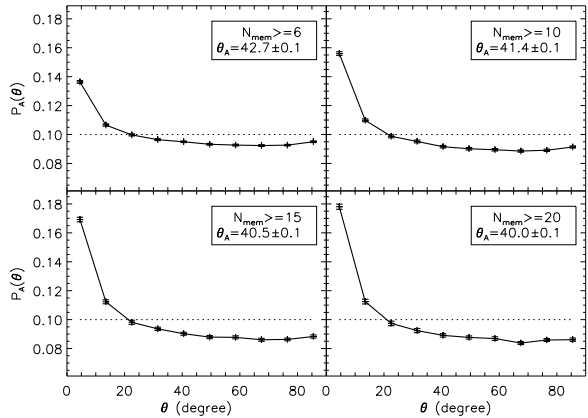}} 
\caption{Dependence of BSG alignment on group richness in SAM.}
 \label{samrich}
\end{figure}

\begin{figure}
\resizebox{\hsize}{!}{\includegraphics{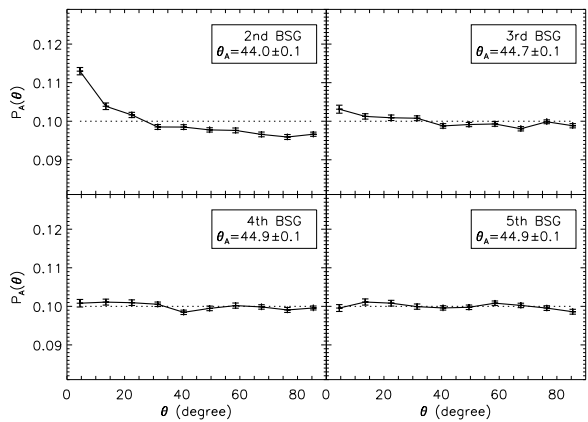}}
\caption{Alignment for other satellites in SAM.}
 \label{samsat}
\end{figure}

  Finally we show in FIGs. \ref{samrich} and \ref{samsat} the richness
  dependence of the BSGs and the alignment signals of other satellite
  galaxies. All these results are quite similar to those obtained from the
  SDSS observations, except that the second brightest satellite galaxies in
  SAM still show some weak major-axis alignment.

\subsection{BSG Merger Time}

\begin{figure}
\resizebox{\hsize}{!}{\includegraphics{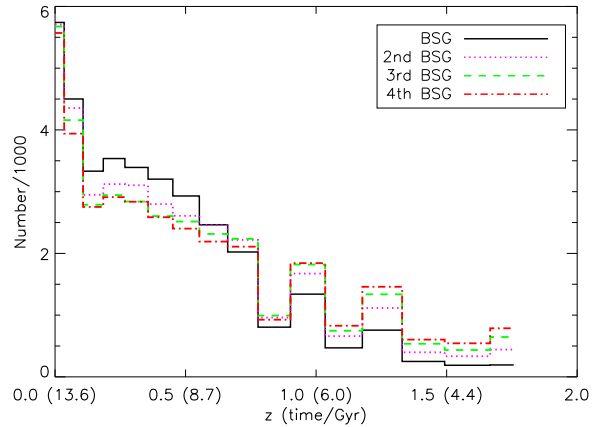}}
\caption{The distribution of merger time of BSGs and other satellites.  The
  black solid line shows the distribution of number of BSGs which merged
  into groups with respect to redshift or time of universe.  The magenta
  dotted, green dashed and red dot-dashed lines are results for the second,
  third and fourth brightest satellite galaxies, respectively. The average
  time when mergers happened are 9.87, 9.58, 9.37 and 9.23 Gyrs for the BSGs,
  second, third and fourth BSGs respectively. } \label{zdist}
\end{figure}

Now we  explore the time  when  BSGs merged  into the host  halos.  We
select all  FOF halos with mass  larger than $10^{13}\smh$ at  $z=0$.  We look
back the progenitors  of BCG and BSG  in each FOF halo along  the merger tree,
until we find the  snapshot in which the progenitors of BCG  and BSG belong to
different halos  for the first time. Then  we define this snapshot  as the BSG
merger snapshot. We also get the merger time for other satellites (the second,
third and  fourth brightest in r  band magnitude) using the  same method.  The
distributions of  merger time for the  BSGs and other satellites  are shown in
FIG.~\ref{zdist}. Here  one can  see that  there are more  BSGs merged  at low
redshifts than other satellites. The average time when BSGs merged is 9.87
Gyrs. While for the second, third and fourth BSGs, the average merger time are
9.58, 9.37 and 9.23 Gyrs  respectively. These support the conjecture that BSGs
merged into the groups  statistically more recently than other satellites,
although the  signal is weak.  Furthermore, $80.7\%$ BSGs in  their progenitor
halos are BCGs before they merged with other halos.

\begin{figure}
\resizebox{\hsize}{!}{\includegraphics{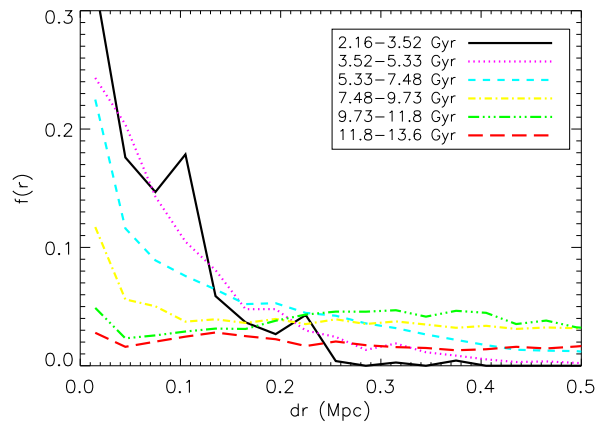}}
\caption{Distribution of distances between BCG  and BSG at $z=0$ in subsamples
  of groups whose BSGs merged in different time.  The black solid line are
  for groups  whose BSGs  merged at  the time  between 2.16 to  3.52 Gyrs,
  while magenta dotted line for BSG merger time between 3.53 to 5.33 Gyrs, and
  so on.}
\label{dz}
\end{figure}

Finally, we check the scenario that  earlier merged BSGs are closer to BCGs at
redshift $z=0$, by dividing the SAM catalog into 6 subsamples according to the
time  when  BSGs merged  and  calculating  the distribution  function  of
distances between BCGs and BSGs at  $z=0$.  As shown in FIG.~\ref{dz}, most of
earlier merged  BSGs are located  within 0.2 Mpc  to their BCGs. While  only a
tiny fraction  of later merged BSGs are  located within 0.2 Mpc  to their BCGs
and most of them are far from the BCGs.

\section{Summary and Discussions}   \label{conclusion}

In this work,  we studied the alignment of BSG-BCG  connection line with group
major axis  using the SDSS  DR4 group catalog  constructed by Y07 and  the SAM
constructed  by  \cite{2011MNRAS.413..101G}.    We  found  good  agreement  of
major-axis alignment  of BSGs  in observations and  SAM.  We also  studied the
distribution of  merger time for  BSGs and other  satellites in SAM  and found
positive evidence to support the scenario  that BSGs merged into their host
groups more recently than other satellites,  and preferentially along the
  filaments.  The main results are summarized as following:

\begin{itemize}
\item A significant group major-axis alignment of BSG is detected.  The
  alignment angle is $43\degree.0\pm 0.4$. The confidence level of BSG
  alignment detection reaches $5\sigma$, owing to the large group
  catalog. BSGs in more massive groups and richer groups show stronger
  alignment with the major axis of their host groups. Furthermore, the
    BSG alignment around blue BCGs is slightly stronger than that around red
    BCGs. And red BSGs have much stronger major-axis alignment than blue
    BSGs.

It is worth to distinguish our BSG alignment from other satellite alignment
measurements. Most of them reflect the global distribution of all satellites
along the major or minor axes of groups, while we measure the preference of
BSG distribution along the major or minor axes relative to the other
satellites.

\item Satellites other than BSGs do not show any significant alignment. This
  may indicate that the BSGs are quite different from other satellites in the
  formation history of groups.

\item In SAM, BSGs are found to have major-axis alignment with similar
  strength as in observations. The alignment show strong dependence on group
  mass and richness in the same way as observations, except that this
  dependence in SAM is stronger.

\item Discrepancy also exists in that the SAM shows opposite BSG color
  dependence, which is most probably induced by the fact that SAM does not
  contain any large scale environment ingredient.

\item BSGs in SAM are found to be merged into the host halos
  statistically more recently. And most of the BSGs ($80\%$) were BCGs before
  their merger and earlier merged BSGs are closer to their BCGs at
  redshift $z=0$.
\end{itemize}

The detected alignment of BSGs along the group major axes and no alignment for
other satellites can be understood if the two popular scenarios of galaxy
formation are {\bf both} true. The first scenario is that the accretion of satellites are
preferentially along the filamentary structures surrounding the dark matter
halos of host galaxies.  The second one is that most massive satellites are
accreted into the host halos more recently and so they have experienced less
phase-mixing and relaxation and better preserved the memory of their accretion
history. Using SAM galaxy catalog, we find positive evidence to support the
second scenario. To fully understand the detected alignment of BSGs, detailed
analysis of the accretion history of BSGs in SAM are necessary, which is
beyond the scope of this paper. However, although is opposite in SAM, the
  BSG color dependence in observation might indicate the first scenario as
  well.

\section*{Acknowledgments}

We thank the referee for the comments and suggestions, which helped to improve
the paper. We thank Qi Guo for her kind help of using SAM data and STAR
cluster and giving us so many good suggestions and comments. This work is
supported by the National Science Foundation of China (NSFC, Nos. 11073024,
11121062, 11233005). XC also acknowledges the support from the John Templeton
Foundation. Y.G.W.  acknowledges the support by the National Science
Foundation of China (Grant No. Y011061001 and No. Y122071001). XW is supported
by Gordon and Betty Moore Foundations.

%

\end{document}